\documentclass[reqno,12pt]{article}
\usepackage{amsmath}
\usepackage{bm}
\usepackage{a4wide,amssymb}
\usepackage{graphicx,xcolor}
\usepackage{epstopdf}
\usepackage{bbm}
\usepackage[normalem]{ulem}

\newcommand{\var}{\varepsilon}
\newcommand{\e}{\varepsilon}

\def\be{\begin{equation}}
\def\ee{\end{equation}}
\def\bea{\begin{eqnarray}}
\def\eea{\end{eqnarray}}

\begin{document}

\begin{titlepage}

\begin{center} 
{\bf \Large{Some Considerations on the Fluid-Dynamical Limit of Particle Systems}}

\vspace{1cm}
{\large M. Pulvirenti and S. Simonella}

\vspace{0.5cm}
{\scshape {\small Dipartimento di Matematica, Universit\`a di Roma La Sapienza\\ 
Piazzale Aldo Moro 5, 00185 Roma -- Italy}}
\end{center}

\vspace{0.5cm}
\noindent {\bf ABSTRACT.} \;
In this note we review and compare mathematical procedures aimed at rigorously deriving fluid–dynamic descriptions of classical particle systems in the large-scale limit of Hamiltonian dynamics. Despite the longstanding nature of this program, its fundamental problems remain open, with only a few notable exceptions that we briefly highlight.

\vspace{1cm}
\begin{center}
{\bf CONTENTS}
\end{center}

\vspace{4mm}

1.\ \ \ \ \ Introduction \dotfill \pageref{sec:intro}

\vspace{3mm}

2.\ \ \ \ \ How to derive the Euler equations from particle systems \dotfill \pageref{sec:EE}

\vspace{3mm}

3.\ \ \ \ \ The Boltzmann equation and the low-density limit \dotfill \pageref{sec:results}

\vspace{3mm} 

4.\ \ \ \ \ The hydrodynamical behaviour of the Boltzmann equation \dotfill \pageref{sec:ste}

\vspace{3mm}

5.\ \ \ \ \ Concluding remarks \dotfill \pageref{sec:tree}

\vspace{3mm}

References \dotfill \pageref{sec:bib}

\thispagestyle{empty}
\end{titlepage}


\section{Introduction} \label{sec:intro}
\setcounter{equation}{0}    
\def\theequation{1.\arabic{equation}}

In this note we discuss the problem of deriving macroscopic fluid and gas equations, starting from a microscopic model consisting of a large number of interacting particles which evolve according to Newton’s laws. Our interest is in the genuinely dynamical problem: how time evolution at the macroscopic scale emerges from the underlying Hamiltonian dynamics of many-particle systems of physical relevance. Apart from a very short comment in Section~5, we restrict our attention to nonlinear models. Throughout, we disregard quantum and relativistic effects. The analysis is discursive rather than technical: our aim is to orient the reader within the existing literature, to recall some basic conceptual issues which are not always stressed in recent works, and to outline the rigorous approaches to the large scale limit problem. 

Our starting point are the Euler equations (EE) for a  compressible fluid. The basic fields are $\rho=\rho(x,t)$, $u=u(x,t)$ and $E=E(x,t)$, representing density, velocity and total energy at the macroscopic point $x$ and time $t$. A point $x$ is understood as a small region containing many particles, so these fields are local averages.

If $\Lambda$ is a macroscopic volume, then
$$
\int_\Lambda \rho =M_\Lambda, \qquad  \int_\Lambda \rho u =P_\Lambda,  \qquad \int_\Lambda \rho E =E_\Lambda
$$
denote the mass, momentum and energy of the fluid in $\Lambda$.

One convenient divergence-form of the EE is
$$
\partial_t \rho +\nabla\!\cdot (\rho u)=0,
$$
$$
\partial_t (\rho u)+\nabla\!\cdot (\rho u \otimes u)+\nabla p=0,
$$
$$
\partial_t E +\nabla\!\cdot \big[(E+p)u\big]=0.
$$
There are many equivalent formulations; we use this one because it exhibits conservation in divergence form. We remind that these equations are time-reversible i.e.\,invariant under change of the signs of $t$ and $u$.

Here $p=p(x,t)$ is the (a priori unknown) pressure. The above identities are purely kinematic; physics enters through the constitutive relation $p=p(\rho,T)$ (with $T$ the temperature) dictated by the interaction potential and local thermodynamic equilibrium: locally, a large particle system is assumed to be at thermal equilibrium.

In what follows we consider smooth solutions of the EE, thus avoiding discontinuities produced by shocks and boundary effects, which would require a substantially more involved analysis.

There are two (at least formal) routes to the EE. The first, arguably the most natural, rescales space and time in the Newtonian dynamics, moving from microscopic to macroscopic variables by the same factor $\varepsilon$, while increasing the number of particles according to $N=\varepsilon^{-3}$ (in three dimensions). Considering the empirical distributions of mass, momentum and energy (and the corresponding currents), one obtains five continuity identities, which close into the EE once local thermal equilibrium is assumed, yielding the thermodynamic relation $p=p(\rho,T)$. This uses the virial theorem to identify mechanical and thermodynamic pressure. This approach is not yet rigorous because of the lack of satisfactory ergodic properties for mechanical systems - a subtle and difficult issue.

The second route (leading to a different set of EE, as explained below) proceeds through the Boltzmann equation (BE)
$$
(\partial_t+v\cdot \nabla_x)f=\lambda\, Q(f,f),
$$
for the one-particle distribution function $f$. Here $\lambda$ is a parameter (fixed for the moment) and $Q$ is Boltzmann's bilinear collision operator.

The transition from Newton’s laws to the BE is justified under the Boltzmann–Grad (low-density) limit, in which space and time are rescaled as above, but $N=\varepsilon^{-2}$, so that only a finite number of collisions occur over a finite (macroscopic) time interval. The variables in which the BE holds are therefore \emph{mesoscopic}, intermediate between micro and macro. If we further rescale time so that $\lambda=\varepsilon^{-1}$, we increase the number of collisions per unit time and can pass (pre-shock) to hydrodynamics. This step can be made rigorous; however, it bypasses the ergodic problem for particle systems, and the resulting EE correspond to the perfect gas law $p=\rho T$: hence they differ from those obtained by the direct hydrodynamic scaling where the equation of state depends on the interaction.

\bigskip
\bigskip
\bigskip
\bigskip
\includegraphics[width=0.9\textwidth]{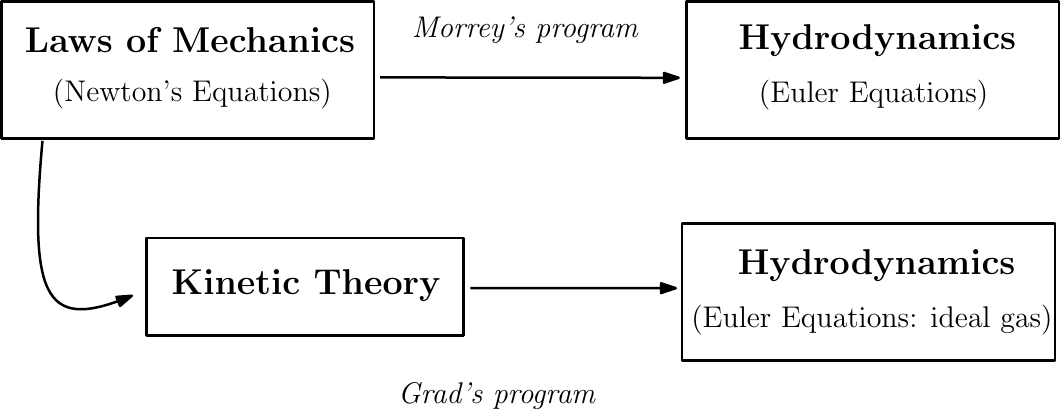}
\bigskip
\bigskip
\bigskip
\bigskip

The figure is meant to summarize the abovementioned regimes. The sets of EE obtained by the two methods are disjoint because the equations of state differ.  The upper and lower paths in the diagram are here attributed to Morrey and to Grad, respectively. These authors were arguably the first to explain that one should connect one class of models to another with a precise mathematical procedure, namely the scaling limit; see \cite{Mo55} and \cite{Gr58}.
Note that, when considering a rarefied gas, the number of collisions per unit time is diverging and the hydrodynamical description obtained via the BE may be more appropriate than the kinetic description.


In the remainder of this note we examine the two complementary approaches. The first is the direct hydrodynamic scaling of the particle system (Section 2), where local thermodynamic equilibrium and the virial theorem lead to the Euler equations with an interaction–dependent pressure law. The second proceeds through the Boltzmann equation (Section 3): under the Boltzmann–Grad (low-density) limit one obtains a kinetic description whose hydrodynamic (pre-shock) limit yields the Euler equations for a perfect gas (Section 4). We compare the scope and limitations of these two derivations, highlighting the conceptual differences between their scaling regimes, and conclude with a brief discussion of extensions toward the Navier-Stokes regime and their relation to Hilbert’s sixth problem (Section 5).

\section{How to derive the Euler equations from particle systems} \label{sec:EE}
\setcounter{equation}{0}    
\def\theequation{2.\arabic{equation}}

In this section we outline a strategy to derive the EE directly from a microscopic system. Our arguments are heuristic; a complete proof remains open and involves hard issues of ergodic theory.

Consider $N$ particles of unit mass moving in the three-dimensional torus $\bar T$. The torus assumption is a drastic simplification whose sole purpose is to avoid boundary effects.
Positions, velocities and time are denoted by $q_1,\dots,q_N$, $v_1,\dots,v_N$, $\tau$. The equations of motion are
$$
\frac{dq_i}{d\tau}(\tau)=v_i(\tau), \qquad 
\frac{dv_i}{d\tau}(\tau)= - \sum_{j\ne i} \nabla 
\phi(|q_i(\tau) -q_j(\tau)|),
$$
with a smooth interaction potential $\phi$. Here $N$ is large and diverging. Introduce a small parameter $\varepsilon$ such that 
$$
N \to \infty, \qquad \varepsilon \to 0, \qquad \varepsilon^3 N =1.
$$

We consider configurations that are almost constant on an intermediate scale, smaller than $\varepsilon^{-1}$. We therefore employ the scaling
$$
x_i=\varepsilon q_i, \qquad t=\varepsilon \tau,
$$
and rewrite the dynamics in the new variables (with the limit $\varepsilon \to 0$ in mind):
$$
\frac{dx_i}{dt}(t)=v_i(t), \qquad 
\frac{dv_i}{dt}(t)= - \frac{1}{\varepsilon} \sum_{j\ne i} \nabla 
\phi\!\left( \frac{|x_i(t) -x_j(t)|}{\varepsilon} \right),
$$
where $x\in T_1$, the unit torus (we assume $\bar T=T_\varepsilon$, of side $\varepsilon^{-1}$). The time scaling reflects the expectation that local thermal equilibrium is reached over a very long microscopic time.

Introduce the empirical densities of mass, momentum and energy:
\be
\nu_0(x,t)=\frac{1}{N}\sum_{i=1}^N \delta(x-x_i(t)),
\label{emp1}
\ee
\be
\nu_\alpha(x,t)=\frac{1}{N}\sum_{i=1}^N \delta(x-x_i(t))\,v_i^\alpha(t), \qquad \alpha=1,2,3,
\label{emp2}
\ee
\be
\nu_{4}(x,t)=\frac{1}{N}\sum_{i=1}^N \delta(x-x_i(t)) \left[\frac{1}{2}
|v_i(t)|^2+\frac{1}{2}\sum_{j\ne
i}\phi(\varepsilon^{-1}|x_i(t)-x_j(t)|)\right],
\label{emp3}
\ee
where $v^\alpha$ are the components of $v$.

We have the continuity equations
\be
\partial_t \nu_\alpha (x,t) =- \nabla \cdot j_\alpha (x,t), \qquad \alpha=0,1,2,3,4,
\label{evemp}
\ee
with the empirical currents of mass, momentum and energy:
\be
j_0 (x,t)=\frac{1}{N} \sum_{i=1}^N \delta(x_i(t) -x)\, v_i(t),
\label{curr1}
\ee
\be
j_\alpha (x,t)= \frac{1}{N} \sum_{i=1}^N \delta(x_i(t) -x) \left[
v_i^\alpha(t) v_i(t) - \frac{1}{2}\sum_{j=1}^N
 \partial_\alpha \phi  \!\left( \frac {x_i(t)  -x_j(t)} {\varepsilon} \right) \left( \frac { x_i(t)  -x_j(t)} {\varepsilon}\right) \right] +O_\varepsilon, 
 \label{curr2}
\ee
for $\alpha=1,2,3$, and
\be
j_4 (x,t)= \frac{1}{N} \sum_{i=1}^N \delta(x_i(t) -x)\, (A_i(t)+B_i(t)) +O_\varepsilon,
\label{curr3}
\ee
where 
$$
A_i(t)=v_i (t)\, \frac{1}{2} \Big(|v_i(t)|^2+ \sum_{j\ne i}\phi(\varepsilon^{-1}|x_i(t)-x_j(t)|)\Big),
$$
$$
B_i^\beta(t)= \frac{1}{2}\sum_{j=1}^N\sum_{\gamma=1}^3 \Psi_{\gamma,\beta} \big(\varepsilon^{-1}(x_i(t) -x_j(t))\big)\, \frac{1}{2}\big[ v^\gamma_i(t) + 
v^\gamma_j(t)\big],\ \ \ \ \ \beta = 1,2,3
$$
with
$$
\Psi_{\gamma,\beta}(z)= -z_\beta(\partial_\gamma \phi)(z).
$$
Here $O_\varepsilon$ denotes an error vanishing with $\varepsilon$.
Equation~\eqref{curr1} is immediate. To show~\eqref{curr2}, set, for a test function $g$,
$$
(g, \nu_\alpha)=\frac{1}{N} \sum_{i=1}^N v_i^\alpha (t)\, g(x_i(t)).
$$ 
Then 
$$
\frac{d}{dt} (g,\nu _\alpha)= (\nabla_x g, j_\alpha)= \frac{1}{N}\sum_{i=1}^N
\sum_{\beta=1}^3(\partial_\beta g)(x_i)\, v_i^\alpha v_i^\beta-\frac{1}{N}\sum_{i=1}^N g(x_i)\,\varepsilon^{-1}\sum_{j\ne i} (\partial_\alpha
\phi)(\varepsilon^{-1}(x_i-x_j)).
$$
By symmetry of the potential, the second term can be written as
$$ 
-\frac{1}{N}\sum_{i=1}^N g(x_i)\,\varepsilon^{-1}\sum_{j\ne i} (\partial_\alpha
\phi) (\varepsilon^{-1}(x_i-x_j))=-\frac{1}{N}\sum_{i,j=1}^N\frac{1}{2}\big(g(x_i)-g(x_j)\big)\,\varepsilon^{-1}(\partial_\alpha \phi)(\varepsilon^{-1}(x_i-x_j)).
$$
Finally,  since $|x_i-x_j| \leq C \var$
$$
\varepsilon^{-1}\big(g(x_i)-g(x_j)\big)= \varepsilon^{-1}\sum_{\beta=1}^d (\partial_\beta g)(x_j)\,(x_i-x_j)
{\color{black}
+ O_\varepsilon\,}
$$ 
from which we deduce
$$
\frac{d}{dt} (g,\nu _\alpha)= \sum_{\beta = 1}^3 (\partial_{\beta} g, j_\alpha^\beta)\;.
$$
Eq.\;\eqref{curr3} is obtained similarly.

Equations~\eqref{evemp} are purely mechanical identities, direct consequences of the equations of motion. To obtain closed equations for the averages of the empirical distributions one must assume a suitable statistical state. Here the local equilibrium hypothesis enters: at each macroscopic point there corresponds a large particle system believed to have reached thermal equilibrium after a very long microscopic time. We therefore consider, as a good approximation, the local 
{\color{black}  grand-canonical Gibbs local equilibrium measure, whose restriction to the $N$-particle phase space, $N\geq 0$, is:} 
\be
P^N (x_1,v_1,\dots,x_N,v_N, t)=Z^{-1}\prod_{i=1}^N
\exp\Big[\mu(x_i,t)-\frac{1}{2T(x_i,t)}\Big((v_i-u(x_i,t))^2 +\sum_{j\neq
i}\phi(\varepsilon^{-1}|x_i-x_j|)\Big)\Big],
\label{Gibbs}
\ee
where $\mu(x,t)$, $u(x,t)$ and $T(x,t)$ are the chemical potential, mean velocity and temperature. $Z$ is the normalization.

Taking expectations of~\eqref{evemp} with respect to~\eqref{Gibbs}, and using the virial theorem 
relating thermodynamic and kinetic pressure at equilibrium, we arrive in the limit $N = \varepsilon^{-3} \to \infty$ at
$$
\partial_t (\rho u)+\nabla \cdot \big(\rho\, u\otimes u\big)+\nabla p=0,
$$
and the corresponding energy equation follows analogously.

A rigorous control of the ergodic properties needed to justify this program is still missing. 

The problem discussed in this section was formulated and partially addressed by Morrey~\cite{Mo55}. Fundamental contributions - explicit introduction of the scaling and the virial theorem - are~\cite{DIPP84} and~\cite{Pr75}; see also~\cite{Sp91,EP04} for reviews. 

A full mathematical result for particle systems has been achieved only in presence of a stochastic noise \cite{OVY93}.

\section{The Boltzmann equation and the low-density limit} \label{sec:results}
\setcounter{equation}{0}    
\def\theequation{3.\arabic{equation}}

In 1872, Boltzmann~\cite{Bo64}, following ideas of Maxwell~\cite{Ma67}, proposed the Boltzmann equation (BE) to describe the time evolution of a rarefied gas. His justification from Newtonian mechanics was subtle and long debated. A remarkable novelty is the emergence of macroscopic irreversibility (approach to equilibrium) from time-reversible particle dynamics. For a historical and bibliographical discussion, see~\cite{Ce06}.

Much later (1949), Grad~\cite{Gr58} predicted the validity of the BE in the limit
$$
N \to \infty, \qquad \varepsilon \to 0, \qquad \varepsilon^2 N =1,
$$
for a gas of $N$ hard spheres of diameter $\varepsilon$: the \emph{low-density} (Boltzmann-Grad) limit. Compare this with the hydrodynamic scaling discussed earlier: rescaling space and time as before but taking $N\sim \varepsilon^{-2}$ (rather than $\varepsilon^{-3}$) captures a regime with a finite number of collisions over finite macroscopic times.

The unknown $f(x,v)$ is now the one-particle probability density; all particles are identical, so the others share the same distribution. Strictly speaking, $f(x,v)\,dx\,dv$ is a probability, distinct from the empirical fraction of particles in a phase-space cell, but the two coincide asymptotically via a law of large numbers as $N\to\infty$.

For hard spheres, the BE reads
$$
(\partial_t+v\cdot \nabla_x)f=Q(f,f),
$$
with collision operator
$$
 Q(f,f) (x,v)=\int dv_2 \int_{S_+} dn \; (v-v_2)\cdot n \;\big[f (x,v') 
f(x,v'_2)-f (x,v)\,f(x,v_2)\big].
$$
Here $(v',v'_2)$ are the post-collisional velocities determined by conservation of energy and momentum; $S_+$ is the half-sphere $\{n:\,(v-v_2)\cdot n\ge 0\}$.

Let $W^N$ be the $N$-particle probability density and $f_j$ its $j$-particle marginals. The BBGKY hierarchy reads
\be
\partial_t f^N_j+\mathcal{L}_j^\varepsilon f^N_j= \varepsilon ^2 (N-j) \, C^\varepsilon_{j,j+1}\,f^N_{j+1},\quad  j<N,
\label{hie}
\ee
where $\mathcal{L}_j^\varepsilon$ is the Liouville operator for $j$ hard spheres and
$$
C^\varepsilon_{j,j+1}f^N_{j+1}(x_1,v_1,\dots ,x_j,v_j)=
 \sum_{k=1}^{j}\int dn \int dv_{j+1} \;
 n \cdot (v_k-v_{j+1}) \;
 f^N_{j+1}(x_1,\dots ,x_j,v_1,\dots ,v_j, x_{j}-\varepsilon n,v_{j+1}).
$$
Solving~\eqref{hie} leads to the perturbative expansion
\begin{align*}
f_j^N(t)=
\sum_{m\ge 0}&\varepsilon^{2m} (N-j)(N-j-1)\cdots (N-j-m+1) 
\int_0^t dt_1 \int_0^{t_1} dt_2 \cdots \int_0^{t_{m-1}} dt_m \\
&\times U^\varepsilon(t-t_1)C_{j,j+1}^\varepsilon \cdots
U^\varepsilon(t_{m-1}-t_m)C_{j+m-1,j+m}^\varepsilon U^\varepsilon(t_m)\,f^N_{0,m+j},
\end{align*}
where $f^N_{0,j}$ are the initial marginals, bounded by $C^j f_0^j$ by assumption (for some $C>0$), and
$$
U^\varepsilon (t)g_j(X_j,V_j) =e^{-\mathcal{L}^\varepsilon_j
t}g_j(X_j,V_j)=g_j(\Phi_j^{-t}(X_j,V_j)),
$$
with $\Phi_j^{t}$ the $j$-particle Hamiltonian flow.

For fixed $j,m$, 
assuming, for simplicity, the velocities uniformly bounded, one estimates the $m$-th term by
$$
\|f_0 \|_{L^\infty}^{m+j}\,C^{j+m} \, \frac{j(j+1)\cdots (j+m-1)}{m!}\, t^m,
$$
and since
$$
\frac{j(j+1)\cdots (j+m-1)}{m!} \le e^{j+m}
$$
the series converges for small $t$, uniformly in $N$; a similar bound holds for the corresponding expansion built from solutions of the BE. Term-by-term convergence then yields convergence of solutions.


This is the strategy devised by Lanford \cite{La75}, who proved the following. 

\emph{Let $f_0$ be a continuous probability density, bounded by a Maxwellian in velocities. Let $f_j^\varepsilon$ solve the BBGKY hierarchy with suitably bounded initial data, and converging to a product of functions $f_0$ in a strong sense. Then, in the Boltzmann–Grad limit and for sufficiently small times $t$, $f^\varepsilon_j$ converges a.e.\ to $\prod_{k=1}^j f(x_k,v_k,t)$, where $f$ is the unique (mild) solution of the Boltzmann equation with datum $f_0$.}

\bigskip

The proof consists of (i) improved estimates allowing unbounded velocities via suitable norms, and (ii) a careful term-by-term convergence analysis. The latter is subtle: for instance, for certain lattice-like mechanical models the series can be controlled for short times but term-by-term convergence fails (see Uchiyama~\cite{Uc88}).

Many mathematical studies followed Lanford’s work; see
\cite{Sp91, CIP94, BGSRS23} for comprehensive accounts.

Recently, Deng, Hani and Ma \cite{DHM24,DHM25} proved a significant extension: validity of the BE for arbitrary times, provided a sufficiently smooth solution exists up to that time - thus reducing validity to an existence problem. The proof introduces new ideas that we do not discuss here.

For other interaction models, different scaling limits exist, belonging to the class of kinetic limits \cite{Sp91}; we do not discuss them here. Notably, this includes the Landau and Balescu-Lenard equations, which extend Boltzmann's theory to plasma physics (cf.\,\cite{NSV18,NVW21}).

\section{The hydrodynamical behaviour of the Boltzmann equation} \label{sec:ste}
\setcounter{equation}{0}    
\def\theequation{4.\arabic{equation}}

Although the BE is derived for a rarefied gas, one may increase the collision frequency to reveal hydrodynamic behaviour. This goes back to Chapman, Enskog and Hilbert and leads to a formal (pre-shock) hydrodynamic limit. There is an apparent contradiction: the Boltzmann-Grad scaling yields a bounded number of collisions per unit macroscopic time, whereas hydrodynamics corresponds to many collisions and rapid local equilibration. In fact, this limit is quite different in nature from the hydrodynamic limit (Section 2) as the starting point is a continuum picture, and time-irreversible.

Multiplying the BE by by $\chi_\alpha=(1,v_1,v_2,v_3,v^2)$ and integrating over $v$ gives
$$
\partial_t \xi^\alpha_f +\nabla_x\cdot \zeta^\alpha_f=0,\quad \alpha=0,\dots,4,
$$
where
$$
\xi^0_f=\rho_f=\int_{\mathbb{R}^3} f\, dv,\qquad
\xi^\alpha_f=\rho_f\, u_{f}^\alpha=\int_{\mathbb{R}^3} v^\alpha f\, dv,\ \alpha=1,2,3,
$$
$$
\xi^4_f=\frac{1}{2}\rho_f |u_f|^2 +
\rho_f e_f, \qquad \rho_f e_f=\int_{\mathbb{R}^3}\frac{1}{2} |v-u_f|^2 f\, dv,
$$
and $\rho_f$, $u_f$, $e_f$ are interpreted as mass density, mean velocity and specific internal energy. The currents are
$$
\zeta^0_{f}=\rho_f u_{f},\qquad
 \zeta^\alpha_{f}=\int_{\mathbb{R}^3} v^\alpha v f\, dv,\quad \alpha =1,2,3,
$$
$$ 
\zeta^4_{f}=
\int_{\mathbb{R}^3}\frac{1}{2} |v|^2 v\, f\, dv.
$$
These are conservation laws but not a closed PDE system, since the currents depend on $f$. As in the particle setting, closure follows from local equilibrium under an appropriate scaling.

Let us proceed now formally by rescaling variables as in Section 2: in macroscopic variables (no external forces), we get
$$
(\partial_t +v\cdot \nabla_x)f=\frac{1}{\varepsilon}\, Q(f,f).
$$
Hilbert sought $f$ as a formal power series
$$
f=\sum_{n=0}^\infty \varepsilon^n f_n,
$$
with the relations
$$
Q(f_0,f_0)=0, \qquad 2Q(f_0,f_1)=(\partial_t+v\cdot \nabla_x)f_{0}, 
$$ 
where $Q$ is the symmetrized collision operator, and other obvious recursive relations. Then $f_0$ must be a local Maxwellian
$$
f_0=M_{\rho,u,T}= \frac{\rho}{(2 \pi T)^{3/2}} \exp\!\left(-\frac{| v-u|^2}{2T}\right),
$$
with parameters $(\rho,u,T)$ such that 
$$
\int (1,v,|v|^2)\, (\partial_t+v\cdot \nabla_x) M_{\rho,u,T}\, dv=0.
$$
These turn out to be precisely the Euler equations for a perfect gas:
$$
\partial_t \rho +\nabla_x\!\cdot(\rho u)=0,\qquad
(\partial_t+u\cdot \nabla_x)u+\frac{1}{\rho}\nabla_x P(\rho,T)=0,
$$
$$
\frac{3}{2}\rho\,(\partial_t+u\cdot\nabla_x)T +P(\rho,T)\,\nabla_x\!\cdot u=0,
$$
with the perfect-gas law $P(\rho,T)=\rho T$.
There are many references for the Hilbert expansion and the rigorous derivation of the Euler equations for times before shocks and without boundaries, starting from \cite{N78,Ca80}. {\color{black} See also \cite{EP04} for a review.} 

Combining the results summarized here with the global (conditional) result of \cite{DHM25} one can show direct convergence of the particle dynamics toward solutions of the EE under a suitable scaling; see \cite{TD25}. This derivation of the EE from particle systems is drastically different from the direct derivation of Section \ref{sec:EE}, in both its scaling and its resulting equation of state (perfect gas). It is in essence the Boltzmann-Grad scaling, with a mild enhancement of the collision rate ($N \sim \e^{-2}\log \log \e^{-1}$). Nevertheless, \cite{DHM25} is a landmark result as it is a first derivation of the EE starting from particle systems in 3D.

\section{Concluding remarks} \label{sec:tree}
\setcounter{equation}{0}    
\def\theequation{5.\arabic{equation}}

We have stressed up to now that the Euler equations can be viewed as Newton’s laws for particle systems on a suitable space–time scale, or as a reasonable approximation in the framework of the Boltzmann equation.  However, from a physical viewpoint these equations are inadequate in many practical situations, for instance in the presence of boundaries (see, for example, the well-known d’Alembert paradox). 
Moreover, we have assumed that we are in a regime where solutions of the Euler equations remain smooth; shocks therefore require a more subtle analysis. In other words, there are physical reasons why the pure scaling arguments discussed here must be extended to provide a more accurate description of reality. 

We only briefly mention the incompressible limit, which is not, strictly speaking, a pure scaling procedure of the type that is the main focus of this note. 

We start by analysing the Hilbert (or Chapman–Enskog) expansion including the next-order term. The result is the Euler system with an additional diffusive term, whose coefficient vanishes with $\var$. Thus, in a hyperbolic scaling, the viscosity term vanishes as $\var \to 0$. To obtain a finite viscosity, one can go to longer time scales by setting (for microscopic variables $q,\tau$)
$$
x=\var q , \quad t=\var^2  \tau.
$$ 
For small $\var$, the diffusive term then remains finite, while the inertial part diverges. At this point one can also rescale the velocities, obtaining the incompressible Navier–Stokes equations for the rescaled velocity field. These approaches are based on \cite{KM81}, which inspired a series of papers deriving the incompressible Navier–Stokes equations via the Boltzmann equation: see \cite{EP04} and the references therein.

From a physical viewpoint, the Navier–Stokes are more appropriate in the presence of boundaries or shocks; however, it is not clear whether they arise in this case from a pure scaling. 

Let us turn now to the famous address at the International Congress of Mathematicians in Paris in 1900, where Hilbert
posed 23 problems as a basis for mathematical research in the forthcoming century \cite{H00}. 
Among these, the sixth is titled {\em Mathematical Treatment of the Axioms of Physics} and reads:

{\it The investigations on the foundation of geometry suggest 
to treat in the same manner, by means of axioms, those physical sciences in
which mathematics plays an important part; in the first rank are the theory of
probabilities and mechanics}.

In his comments, Hilbert mentions several physicists and Boltzmann's recent work
and expresses the need for a mathematical approach to scaling limits, starting from fundamental 
particle models in order to obtain macroscopic descriptions of fluids and gases; see \cite{PS16}.

Hilbert’s Sixth Problem 
should be further specified.  We have discussed up to now three types of convergence that can be analysed: 

1) Derivation of the Boltzmann equation from particle systems;

2) Derivation of the Euler and/or Navier-Stokes equations from the Boltzmann equation;

3) Derivation of the Euler and/or Navier-Stokes equations from particle systems.

It is not easy to determine which specific problem Hilbert had in mind in \cite{H00}. In a modern perspective, in any case, it is more natural to interpret Hilbert's sixth problem as a remarkable anticipation of an entire research area rather than as a single, well-posed question. This poses a challenge for the future.

We can only mention two exceptional situations in which a macroscopic limit has been rigorously derived from deterministic many-particle dynamics, without invoking any approximation. The first is the case of the periodic Lorentz gas (or Sinai billiard), where dispersing billiard dynamics at fixed scatterer density is sufficiently chaotic to yield (linear) diffusive behaviour \cite{BS81,Sz24}. A second class of examples arises from integrable one-dimensional systems, starting from the classical hard-rod model, which admit an Euler-scale limit
\cite{BDS83}. More recently, generalised hydrodynamics is being developed for a broader family, both classical and quantum \cite{Sp23}. These cases remain special, but they show that, under strong chaoticity or integrability, a rigorous macroscopic description can indeed emerge directly from Hamiltonian many-body dynamics.

We conclude by recalling that there are also completely different approaches, with wide literature dedicated, in which a given continuum system is approximated by suitable stochastic processes (for instance for numerical simulations). 
We refer the reader to \cite{DeMP91,KL91,Sp91,EP04} for results and further references. 

To summarize, in this short contribution we have emphasized that, in the absence of shocks and boundaries, the hydrodynamic limit of particle systems - under a pure scaling procedure - yields precisely the Euler equations for real fluids. The approaches based on the Boltzmann equation may provide an accurate description only in very special physical regimes, and the resulting Euler equations differ from those obtained directly from particle systems. The Navier-Stokes equations should be viewed as a correction to the Euler picture; nevertheless, under a parabolic scaling (with respect to which they are invariant), the incompressible Navier-Stokes equations describe the evolution of fluctuations around a vanishing velocity field.

\bigskip
{\bf Acknowledgments.} We thank R. Esposito and R. Marra for useful discussions.
Funded by the European Union (ERC CoG KiLiM, 101125162).

\end{document}